# Large Chiroptical Effects in Planar Chiral Metamaterials

Weimin Ye,[1] Xiaodong Yuan,[1] Chucai Guo,[1] Jianfa Zhang,[1] Biao Yang,[2] and Shuang Zhang[2,*]

[1]*College of Optoelectronic Science and Engineering, National University of Defense Technology, Changsha, 410073, China.*

[2]*School of Physics & Astronomy, University of Birmingham, Birmingham, B15 2TT, UK.*

* E-mail: s.zhang@bham.ac.uk

## ABSTRACT

**Chiroptical effects, characterized by different optical responses for left- (LCP) and right- handed circularly polarized light (RCP), are powerful and valuable tools in optics with wide applications in polarization resolved imaging and sensing. Previously observed strong chiroptical effects are limited to metamaterials with complex three-dimensional chiral structures at the sub-wavelength scale. Although asymmetrical transmission of LCP and RCP have been investigated in planar chiral metasurfaces, the observed weak chiroptical effects result from anisotropic Ohmic dissipation of the metal constituents. Here, we demonstrate by theory and proof-of-concept experiments that large difference in transmittances of LCP and RCP can be attained in a single-layer planar chiral metamaterial with a sub-wavelength thickness. Without violating the reciprocity and mirror symmetry, the strong chiroptical effect, independent of dielectric loss, arise from a novel mechanism of multimode interference. The described effect may lead to new gateway towards chiral manipulations of light and chiral optical devices.**

## I. INTRODUCTION

The word chirality comes from the Greek word for the human hand, which plays a very important role in various subjects including mathematics, physics, chemistry and biology. According to the definition of Lord Kelvin, any geometric figure or group of points is chiral if its image in a plane mirror cannot be brought to coincide with itself [1]. In optics, chirality breaks the degeneracy between the two spins of photons. Thus, chiroptical effects [2, 3] are characterized by the different optical responses for the left-handed (LCP) and right-handed circularly polarized light (RCP). One of the typical effects is optical activity, which is the ability to rotate the plane of linearly polarized light about the propagation direction. Another one is circular dichroism measured by the difference in absorptions of LCP and RCP. Extending to lossless systems or focusing on the transmitted light, the chiroptical effect is manifested as the difference in transmittances of LCP and RCP [3], which is also called circular dichroism in transmission (CDT) [4]. Attaining strong chiroptical effects has great significance for applications in polarization resolved detection and imaging. However, optical chirality of natural materials is extremely small [2, 4], leading to negligible optical effect. Benefiting from the advances in nanophotonics and nanofabrication techniques, artificial nanostructured materials, such as photonic crystals and metamaterials, enable manipulation of light on a scale comparable or smaller than the wavelength of incident light [5], which has opened opportunities to produce and tailor chiroptical effects that drastically exceed those of natural materials. Strong CDT has been demonstrated in three-dimensional (3D) spiral [6], bi-chiral [7] and gyroid [8] photonic crystals through presence of stop bands for just one circular polarization. A miniature chiral polarizing beam splitter at near-infrared wavelengths was also realized by gyroid photonic crystals [9]. Introducing noble metals into the nanostructures, broadband circular polarizers at midinfrared wavelengths have been demonstrated by using gold helix photonic metamaterial [10]. Both circular dichroism and optical rotatory dispersion effects at visible wavelengths were observed in DNA-based helical arrangements of achiral metallic nanoparticles [11]. In addition to 3D helical structure, 3D multilayered structures also exhibit chiroptical effects. A strong optical activity effect at the microwave band has been measured in a bilayer [12] and trilayer [13] chiral metamaterials. Asymmetric transmissions of circularly polarized light (CP) and a circular polarizer in the optical band have been realized by cascading two [14], three [15] layers of achiral metasurfaces and twisted metamaterial slabs [16].

In comparison to 3D chiral metamaterials, planar chiral metamaterials (PCMMs) [17, 18], which preserve mirror symmetry in light propagation direction and are chiral only in the two dimensions, possess obvious technological advantage for easy fabrication and on-chip integration. Among them, planar chiral metasurfaces (PCMSs) [4, 17-21] with thickness far less than the wavelength of incident light have attracted great attention. Unfortunately, owing to the mirror symmetry and reciprocity [22, 23], PCMM itself does not exhibit optical activity [17]. The reported optical activity in quasi-two-dimensional planar nanostructures results from the broken mirror symmetry due to the index mismatch between the superstrate and the substrate [17]. Handedness-dependent [4] or anisotropic [20, 21] Ohmic dissipation of the metal constituents is crucial for the observed CDT or asymmetric transmission of CP from PCMSs, which is much weaker than that from 3D chiral metamaterials. To increase the absorption, a plasmonic PCMS was cascaded with homogeneous dielectric spacer and metal backplane to form a trilayer chiral metamaterials [24, 25], which primarily absorbs one-handed CP and reflects CP of the other handedness. Here, we theoretically and experimentally demonstrate that strong chiroptical effect manifested by large CDT can be attained in a single-layer PCMM, which is independent of the dielectric loss. In contrast to CDT based on circular polarization stop bands, CDT in PCMMs is induced by asymmetric conversion of CP due to the restrictions of the reciprocity and mirror symmetry. In other words, the PCMM transmits and meanwhile reverses the spin for the incident light of a particular circular polarization, and reflects and maintains the spin for the opposite circular polarization. Using the scattering matrix method, we verify that the strong chiroptical effect is the consequence of multimode interference. Our work provides a mechanism to enhance the chiroptical responses in the planar structure and may lead to a gateway towards chiral manipulations of light and chiral optical devices.

## II. DESIGN OF THE PLANAR CHIRAL METAMATERIAL

As illustrated in Fig. 1(a), the proposed PCMM consists of single-layer **L**-shaped gold nano-antennas in a square lattice with period *a* and two primitive lattice vectors along *X* and *Y* directions, respectively. The two arms of **L**-shaped nanoantenna, oriented in *X* and *Y* directions, have identical length *L* and thickness *H* but different widths $W_x$ and $W_y$ as shown in Fig. 1(b). The difference between the two widths breaks the mirror symmetry in the planes parallel to Z direction, leading to two-dimensional geometric chirality for the PCMM. Meanwhile, all the nano-antennas are buried in silica ($SiO_2$) to preserve the mirror symmetry of the PCMM in the propagation direction of light. In addition, to exclude the extrinsic chirality [26] appearing at oblique incidence, only the transmission of CP at normal incidence through the PCMM is investigated. Further, the wavelength of the incident light is constrained to be larger than the lattice period to eliminate the diffracted light in the far field.

Focusing on the chiroptical effect in near-infrared wavelengths, we calculate the transmittance and reflectance spectra of RCP and LCP through the PCMM by COMSOL MULTIPHYSICS software. The refractive index of $SiO_2$ is 1.46. The relative permittivity of gold is modeled by Drude model with $\omega_p$=1.37×$10^{16}$ rad $s^{-1}$ and $\gamma_p$= 4.08×$10^{13}$ rad $s^{-1}$. Figure 1(c) shows that within the wavelength interval between 1460nm and 1600nm, the transmittance of LCP is less than 0.015, whereas that of RCP is greater than 0.8. The transmitted light of an incident RCP is almost completely converted to LCP with ellipticity less than -0.81, which results from the fact [in Fig. 1(d)] that the PCMM only supports a high transmission rate from RCP to LCP (denoted by $T_{LR}$). Meanwhile, transmissions from RCP to RCP ($T_{RR}$), LCP to RCP ($T_{RL}$) and LCP to LCP ($T_{LL}$) are nearly inhibited. Reflectance spectra [Fig. 1(e)] show that the reflectance of LCP within the same wavelength interval is greater than 0.9. The reflected light is nearly LCP [Fig. 1(f)] with the ellipticity greater than 0.8. Moreover, the absorbance and circular dichroism of the PCMM at the simulated wavelengths remain less than 0.1 and 0.02, respectively, indicating that the large CDT is independent of the Ohmic dissipation of gold. This loss independence is further confirmed by the simulation of the PCMM made from an ideal metal (see the Supplemental Material [27]).

### III. MULTIMODE INTERFERENCE IN THE PCMM

In order to investigate the origin of the strong chiroptical effect in the PCMM, we utilize the scattering-matrix method [28], which is a powerful tool for studying planar layered structures of infinite transverse dimension, to numerically investigate the coupling between internal and external electromagnetic fields of the PCMM. The first step of the scattering-matrix method is to solve the eigenmodes of the **L**-shaped gold periodic waveguide array. Figure 2(a) and (b) show the real and imaginary parts of effective mode indexes of the four modes supported by the gold periodic waveguide array with the lowest cutoff frequencies. Based on their transverse-electric-field distributions at the wavelength of 1500nm shown in Fig. 2(c)-2(f), we name the four modes Y-arm, X-arm, X-end, Y-end modes, respectively. The Y-arm and X-arm modes in Fig. 2(c) and 2(d) are dispersive waveguide modes [in Fig. 2(a)] and induced mainly by surface charges located along the *Y*-orientation arm and *X*-orientation arm, respectively. As a result of the geometric chirality of the **L**-shaped structure, the two eigenmodes are nondegenerate in a similar way as the TE eigenmodes in the rectangular metallic waveguide with different cutoff frequencies inversely proportional to the widths of waveguide. On the other hand, the X-end and Y-end modes in Fig. 2(e) and 2(f) are weakly dispersive plasmonic modes [in Fig. 2(a)] and induced mainly by the surface charges located at the *X*- and *Y*-oriented end facets of the **L**-shaped structure. Owing to the geometric chirality and couplings among the array of waveguides in the square lattice, electric fields of the four eigenmodes without binding at the surfaces of metallic waveguides exhibit low symmetry and different local distributions at the sub-wavelength scale. Thus, all of them can couple with an incident plane wave.

The scattering matrix of the PCMM is subsequently obtained by numerical simulations for the scattering process schematically described by Fig. 3(a). Utilizing it, we can easily investigate the chiroptical properties of the PCMM with different thicknesses. For a normally incident CP at the wavelength 1500nm, Fig. 3(b) and 3(d) show the transmittance and reflectance of RCP and LCP , respectively, through the PCMM for varying thickness *H*. We show that the CDT and circular dichroism in reflection are very small for the PCMM with the thickness smaller than 200nm, which is consistent with the weak chirality of previously reported PCMS [4, 18, 20, 21]. However, strong chiroptical effects appear when the thickness is near 360nm, with the transmittance and reflectance of RCP reaching 0.94 and 0.01, respectively. In contrast, the transmittance and reflectance of LCP are equal to 0.01 and 0.92, respectively. Note that the strong chiroptical effects of the PCMM are independent of the Ohmic dissipation of the gold. Furthermore, the calculation shows that the PCMM converts RCP into LCP in transmission with an ellipticity of -0.91, and reflects LCP into LCP with an ellipticity of 0.91. Here, the ellipticity of mode $\left(\mathbf{e}_x \pm i\mathbf{e}_y\right)$ is defined to be $\pm 1$. These results from our simplified scattering-matrix model agree well with those of COMSOL simulations denoted by symbols in Fig. 3(c)-(f), confirming that the strong chiroptical effect arises from the interference among the four modes supported by the **L**-shaped gold waveguide array.

It should be noted that each waveguide mode experiences not only reflection but also converts into other modes at the two PCMM/silica interfaces. This interconversion among the modes at the interfaces greatly complicates the optical processes, and, therefore, a simple picture of the chiroptical effects cannot be readily obtained. Nonetheless, we show that the observed strong chiroptical effect is facilitated by the interference of multiple roundtrips across the PCMM, i.e. the Fabry-Pérot resonance, through an analysis based on the transmission and reflection matrix model as illustrated in Fig. 4(a). We show in Fig. 4(b) that the single-pass transmission already shows some chiroptical effects with transmissivity $T_{\text{LR}}^{(0)}$, $T_{\text{RL}}^{(0)}$ and $T_{\text{RR}}^{(0)}$ ($T_{\text{LL}}^{(0)}$) equal to 0.45, 0.22 and 0.06, respectively. By increasing the number of round-trips in the PCMM, the chiroptical effect increases and converges after 15 round-trips, exhibiting strong chiroptical effects with $T_{\text{LR}}$ reaching 0.94, and the other three transmission coefficients $T_{\text{LL}}$, $T_{\text{RR}}$ and $T_{\text{RL}}$ approaching zero. Figure 4(c) shows the evolution of the chiroptical effect in reflection with increasing number of round-trips, which exhibits similar trend and converging behavior as that of transmission.

## IV. EXPERIMENTAL VERIFICATION OF THE LARGE CHIROPTICAL EFFECTS

For experimental verification of strong chiroptical effects of PCMM, we fabricated a sample on a 0.5-mm-thick double-side polished $SiO_2$ substrate. A 5-nm-thick chromium (Cr) adhesion layer is first deposited on the substrate using *e*-beam evaporator to improve the adhesion of gold. A 360-nm-thick gold layer was sequentially deposited using a dc sputter. Next, the focused-ion-beam (FIB) system is used to mill the gold layer to obtain the designed periodic array of **L**-shaped nanoantennas. To mimic nanoantennas buried in $SiO_2$, a silica refractive-index-matching liquid is added on the patterned gold layer. Finally, another $SiO_2$ wafer that is same as the substrate is covered on the sample for the purpose of protection. The footprint of the sample is about $43\times43\mu m^2$. Figures 5(a) and 5(b) show the scanning-electron-microscopy (SEM) images of the fabricated **L**-shaped nanoantennas for a small area across the sample. It is obvious in Fig. 5(a) that the cross sections of fabricated L-shaped gold nano-antennas are trapezoid, which breaks the mirror symmetry of nanoantennas. Thus, the fabricated sample can be considered a 3D chiral structure.

The sample is tested by using a broadband light source. We measure the circular-polarization-resolved transmittance spectra for RCP and LCP at normal incidence on the sample. The solid lines in Fig. 5(c) are measured overall transmittance spectra for RCP and LCP, which is in good agreement with the simulation results (dashed line) based on the geometry of the fabricated sample. Within the wavelength interval between 1420nm and 1480nm, the measured CDT is greater than 0.4, which was much larger than previous reported results with planar chiral structures [20]. Because the mirror symmetry is broken by the trapezoid cross section of **L**-shaped nanoantennas, the transmittance from RCP to RCP is different from that from LCP to LCP in Fig. 5(d). It should be noted that the measured chiroptical effect is far below that of the proposed PCMM structure (Fig. 1(c)) in terms of both the CDT and the operating bandwidth. The reason is that the trapezoid cross sections of **L**-shaped nanoantennas break the invariance of the PCMM along the propagating direction of light, weakening the Four-mode interference effect which plays the key role in the observed chiroptical effects. Therefore, our experiment can serve only the purpose of a proof of concept of this mechanism for generating strong chiroptical effect with planar structures.

## V. CONCLUSION

In summary, we demonstrate that strong chiroptical effects can be attained in a single-layer planar structure by utilizing a PCMM made up of **L**-shaped gold nanoantennas. Without violating the reciprocity and mirror symmetry, the large CDT exhibited by the PCMM arises from a multimode interference mechanism. Benefiting from the fact that the effective mode indexes of waveguide array could be increased by a higher refractive index of the dielectric constituent, the optimized thickness of the proposed PCMM could be further reduced by burying relatively smaller **L**-shaped gold nano-antennas inside high-index materials such as silicon. In comparison with the reported 3D chiral structures fabricated by direct laser writing and 3D multilayered structures fabricated by multi-step lithography and lift-off, the PCMM can be fabricated by one step of etching, which provides technological advantages for fabrication and on-chip integration with micronanophotonic and electronic systems.

## ACKNOWLEDGMENTS

This work is supported by the National Natural Science Foundation of China under Grants No. 11374367 and No. 61404174. We acknowledge the Center of Material Science of National University of Defense Technology for help with the FIB Fabrication. We also thank Wei Xu for help with the COMSOL simulation and Wei Chen for help with fabrication. S. Z. acknowledges the financial support from ERC Consolidator Grant (TOPOLOGICAL) and the Royal Society.

**FIGURES**

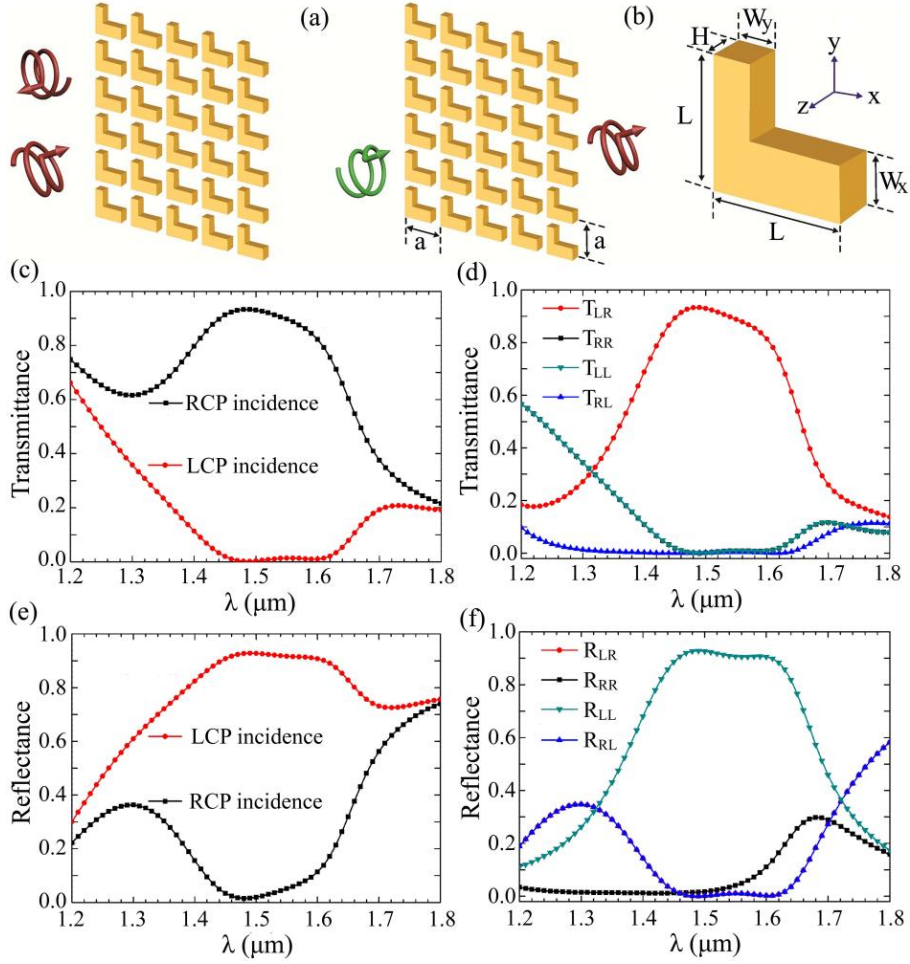

**FIG. 1.** Schematic illustration and theoretical demonstration of the strong chiroptical effect in the planar chiral metamaterial. The proposed PCMM (a) consists of single-layer **L**-shaped gold nanoantennas in a square lattice. Where RCP at normal incidence is converted into LCP when transmitted, LCP is reflected into LCP. The period $a$, thickness $H$, length $L$, widths of two arms $W_x$ and $W_y$ of chiral **L**-shaped gold nanoantenna (b) are 730, 580, 360, 246, and 160nm, respectively. Calculated transmittance (c) and circular-polarization-resolved transmittance (d) spectra for RCP and LCP at normal incidence to the proposed PCMM, respectively. $T_{LR}$ ($T_{LL}$) and $T_{RR}$ ($T_{RL}$) denote the LC- and RC-polarized transmittance of an incident RCP (LCP), and (e), (f) are reflectance and circular-polarization-resolved reflectance spectra. $R_{LR}$ ($R_{LL}$) and $R_{RR}$ ($R_{RL}$) denote the LC- and RC-polarized reflectance of an incident RCP (LCP).

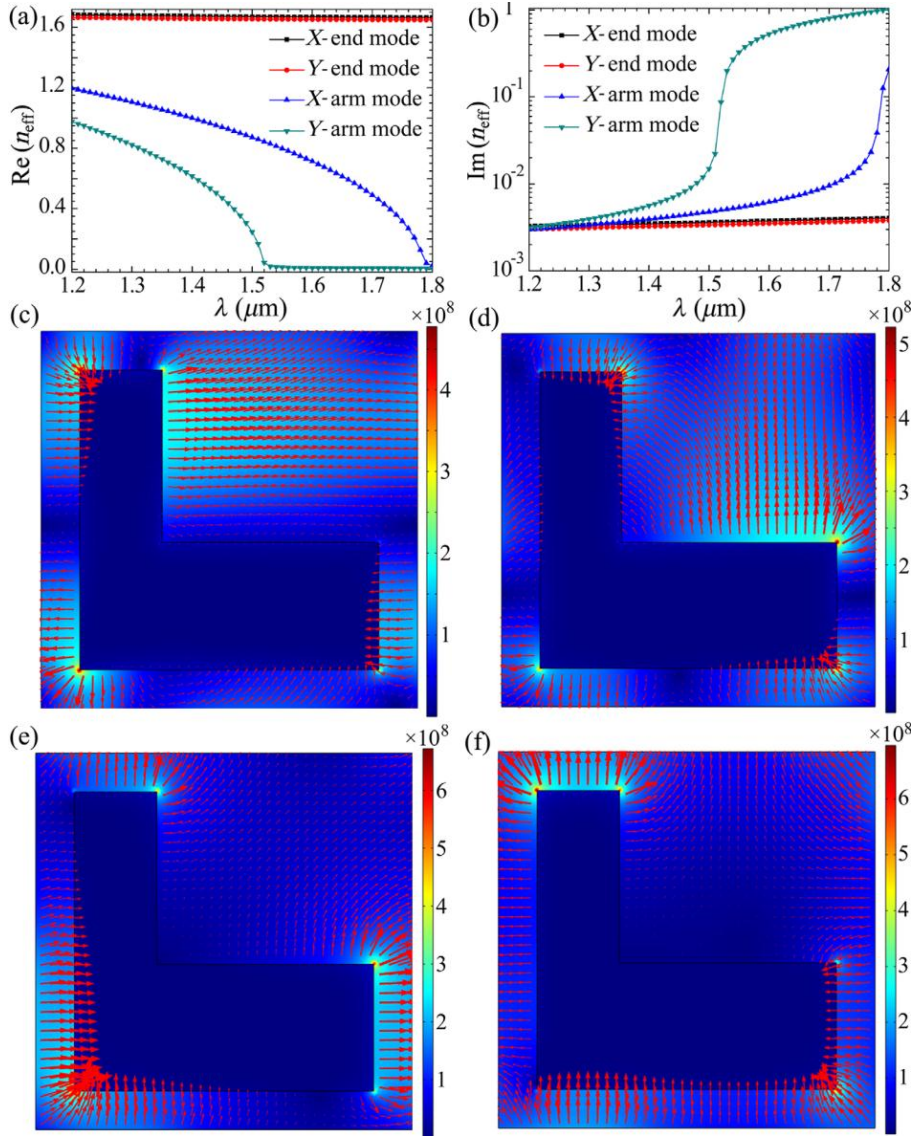

**FIG. 2.** Eigenmodes of **L**-shaped gold periodic waveguide array. Real part (a) and imaginary part (b) of the effective mode index $n_{eff}$ of the four eignmodes supported by **L**-shaped gold periodic waveguide array with the lowest cutoff frequencies denoted as the Y-arm, X-arm, X-end, and Y-end modes according to their transverse-electric-filed distribution (c)-(f) at the wavelength of 1500nm. The small arrows denote the directions of transverse-electric-filed vectors, and the different color hues denote the relative amplitudes of the electric fields.

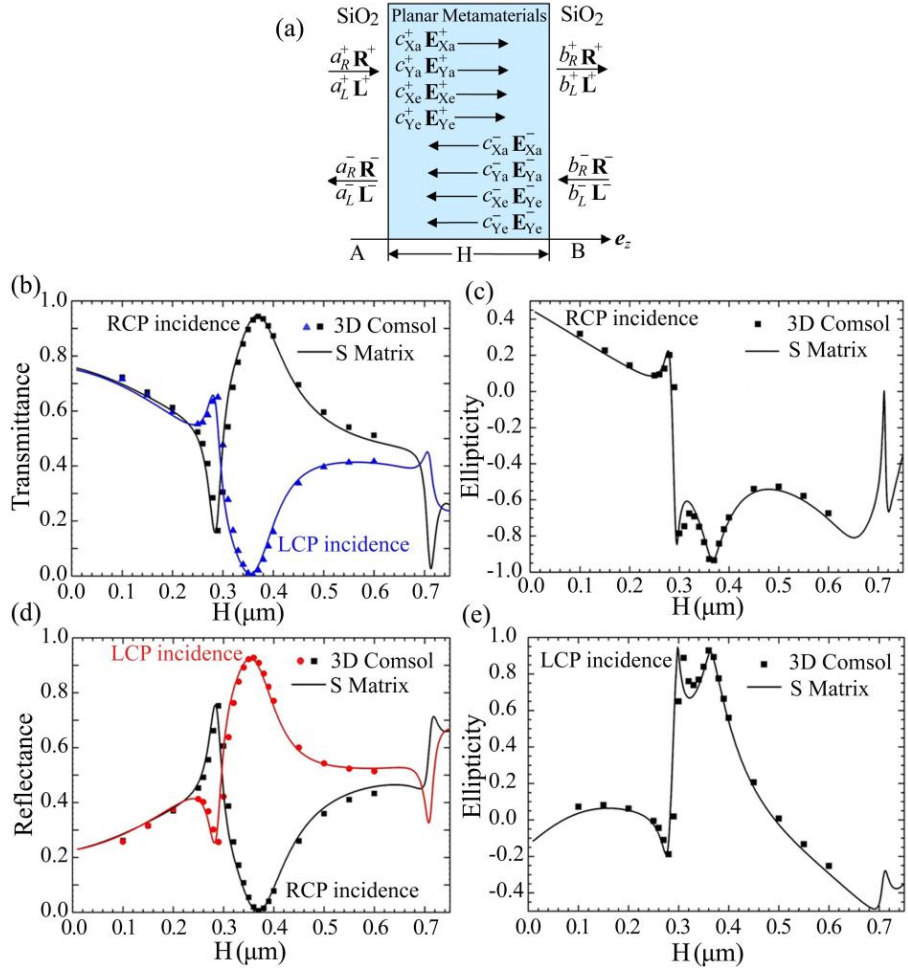

**FIG. 3.** Chiroptical effects in the PCMM with different thickness studied by the scattering-matrix method. (a) Scattering-matrix model of the proposed PCMM. $R^{\pm}$ ($L^{\pm}$) denotes the electric fields of the $\pm z$-directionally propagating RCP and LCP in the semi-infinite $SiO_2$ layer. The electric fields in the PCMM denoted by subscript Xa, Ya, Xe and Ye are that of the four eigenmodes supported by **L**-shaped gold periodic waveguide array shown in Fig. 2 (X-arm, Y-arm, X-end, Y-end, respectively). Transmittance (b) and reflectance (d) for RCP and LCP with the wavelength 1500nm at normal incidence on the PCMM with different thickness $H$. Ellipticity of high-transmittance (c) and high-reflectance (e) circular polarization with different thickness. The lines and symbols (square, triangle, and circle) are results calculated by using the scattering- matrix method and COMSOL simulations, respectively.

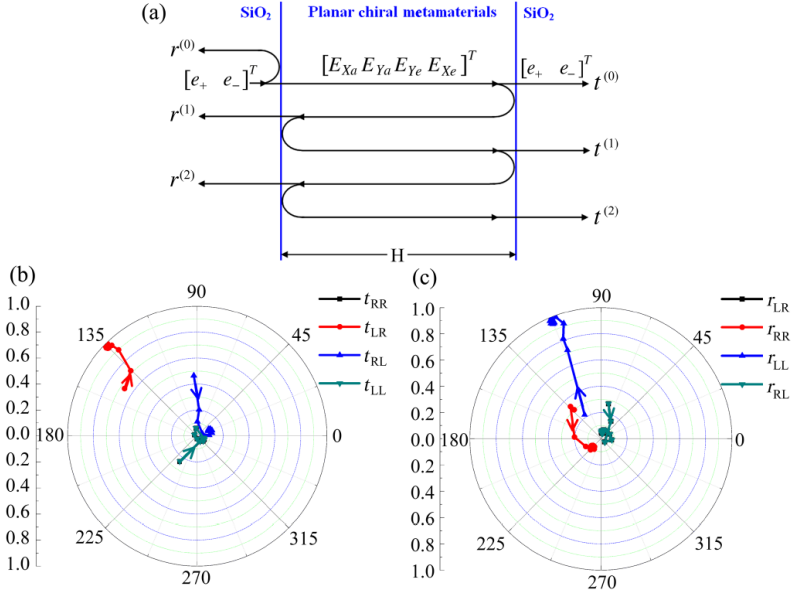

**FIG. 4.** Chiroptical effects in the PCMM studied by the multiple-beam interference model. Transmission and reflection matrix model (a) is used to describe the transmitted and reflected light. The circular-polarization-resolved transmission coefficient $t$ (b) and reflection coefficient $r$ (c) for RCP and LCP with the wavelength 1500nm at normal incidence on the PCMM with 360nm thickness, which varies with the incremental maximum of round-trip from 0 to 20.

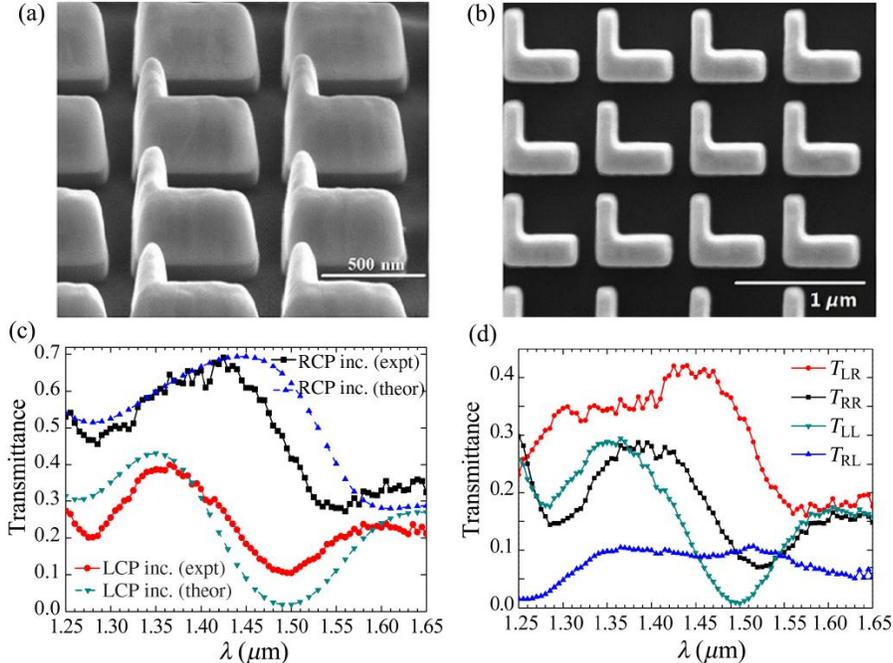

**FIG. 5.** Experimental demonstration of chiroptical effects in the PCMM. SEM images of the fabricated **L**-shaped gold nanoantennas with the sample tilted by $52^0$ (a) and $0^0$ (b), respectively. Measured overall transmittance spectra (c) and circular-polarization-resolved transmittance spectra (d) for RCP and LCP at normal incidence (inc.) on the fabricated PCMM. The dashed lines in (c) are the calculated transmittances of the PCMM consisting of **L**-shaped gold nano-antennas with trapezoid cross sections.